\documentclass[multphys,vecphys]{svmult}

\usepackage{makeidx}         
\usepackage{graphicx}        
\usepackage{multicol}        
\usepackage[bottom]{footmisc}

\makeindex             

\begin{document}

\title*{A study of the magnetic helium variable emission-line star HD\,125823.}
\author{
S.\ Hubrig\inst{1},
N.\ Nesvacil\inst{2},
F.\ Gonz\'alez\inst{3},
B.\ Wolff\inst{4},
I.\ Savanov\inst{5}}
\authorrunning{Hubrig et~al.}
\institute{
European Southern Observatory, Casilla 19001, Santiago, Chile \texttt{shubrig@eso.org},
\and Institut f\"ur Astronomie, Universit\"at Wien, T\"urkenschanzstr.\ 17, 1180 Vienna, Austria \texttt{nicole@jan.astro.univie.ac.at},
\and Complejo Astron\'omico El Leoncito, Casilla 467, 5400 San Juan, Argentina \texttt{fgonzalez@casleo.gov.ar},
\and European Southern Observatory, Karl-Schwarzschild-Str.\ 2, 85748 Garching, Germany \texttt{bwolff@eso.org},
\and Astrophysikalisches Institut Potsdam, An der Sternwarte 16, 14482 Potsdam, Germany \texttt{isavanov@aip.de}}
%
%
\maketitle


The 5.9\,M$_\odot$ star HD\,125823
is a striking helium variable with a period of 8.82\,d, ranging in helium spectral type
from He-strong B2 to B8 (e.g., Norris 1968 \cite{Norris68}).
In fact, HD\,125823 seems to be a transition object between the He-weak and 
He-strong stars with unreddened colors just at the boundary between the two groups
in the UBV color-color diagram. Although high-dispersion studies have been 
carried out in the past by several authors (e.g. Wolff \& Morrison 1974 \cite{WolffMorrison74}), no 
abundance analysis is available.

\begin{figure}
\centering
\includegraphics*[width=0.46\textwidth,clip=]{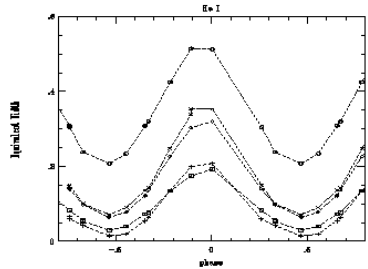}
\caption{
Variations of the equivalent widths of the He~I lines
$\lambda\lambda$4121, 4438, 4713, 5047, and 6678 over the rotation period.
}
\label{hubrig:fig1}       
\end{figure}

\begin{figure}
\centering
\includegraphics*[width=0.46\textwidth,clip=]{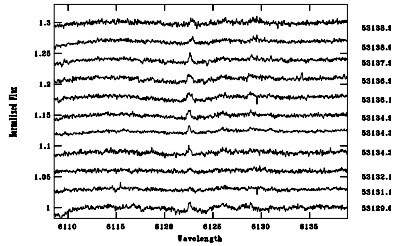}
\caption{
Mn~II emission lines of multiplet 13 in the spectra of 
HD\,125823 obtained on nine consecutive nights with UVES.
}
\label{hubrig:fig2}       
\end{figure}



Recently, we obtained high spectral resolution high signal-to-noise UVES
spectra of HD\,125823 over the rotation period of 8.82\,d 
in order to study the surface distribution of various 
chemical elements.
In Fig.~\ref{hubrig:fig1} we show the behaviour of the He~I lines 
$\lambda\lambda$4121, 4438, 
4713, 5047, and 6678 over the rotation period.
The magnetic field observations reported in the literature (Wolff \& Morrison 
1974 \cite{WolffMorrison74};
Borra et~al.\ 1983 \cite{BorraEtal83}) indicate that the negative
extremum coincides closely in phase with the maximum helium strength. However,
the positive magnetic pole shows a He-deficient cap, so that the two magnetic poles
are associated with very different helium abundances. 

An additional interest to study HD\,125823 comes from the recent detection of 
weak emission lines of various ions (Mn~II, Fe~II, Cr~II, Ti~II, etc.) in optical 
spectra of B-type stars (e.g., Sigut et~al.\ 2000 \cite{SigutEtal00}; Wahlgren \& Hubrig 
2000 \cite{WahlgrenHubrig00};
Castelli \& Hubrig 2004 \cite{CastelliHubrig04}). For the first time we present here 
observational evidence 
for the appearance of emission lines of Mn and Fe
in a variable magnetic star. In Fig.~\ref{hubrig:fig2} we show the behaviour of the emission 
line profiles of the Mn~II multiplet 13.
It is remarkable that the
emission lines of Mn~II are definitely variable, changing their appearance from 
the first night to the last.
To date, explanations of this
phenomenon have been put forward in the context of non-LTE line formation 
and possible fluorescence mechanisms. The qualitative and quantitative assessments 
of the spectra of B type stars suggest 
a possible correlation of the appearance of diverse emission lines with the 
non-magnetic P-Ga and HgMn groups as well as with spectral type. 

From the behaviour of the line profile variations of various elements we can conclude 
that helium is enhanced in 
regions of the stellar surface where silicon and other metals are depleted, and 
helium is depleted in regions where the metals are enhanced. 
A future goal is to derive surface abundance maps for various elements.
Such a study is an important step towards understanding the
effect of magnetic fields on the development of surface chemical
peculiarities in the photospheres of hot stars.


\begin{thebibliography}{}

\bibitem{BorraEtal83}
Borra E.~F., Landstreet J.~D., Thompson I., 1983, ApJS, 53, 151 

\bibitem{CastelliHubrig04}
Castelli F., Hubrig S., 2004, A\&A, 425, 263 

\bibitem{Norris68}
Norris J., 1968, Natur, 219, 1342 

\bibitem{SigutEtal00}
Sigut T.~A.~A., Landstreet J.~D., Shorlin S.~L.~S., 2000, ApJ, 530, L89 

\bibitem{WahlgrenHubrig00}
Wahlgren G.~M., Hubrig S., 2000, A\&A, 362, L13 

\bibitem{WolffMorrison74}
Wolff S.~C., Morrison N.~D., 1974, PASP, 86, 935 

\end{thebibliography}


\end{document}